\documentstyle[12pt]{article}
\begin{document}
\begin{titlepage}

\centerline{{\bf INTERLAYER VORTICES AND EDGE DISLOCATIONS IN HIGH}}
\centerline{{\bf TEMPERATURE SUPERCONDUCTORS}}
\centerline{A.B. Kuklov${\rm ^1}$, A. Krakovsky${\rm ^2}$, and Joseph
L. Birman${\rm ^1}$}
\centerline{\it ${\rm ^1}$Department of Physics, The City College of the
City University of New York,}
\centerline{\it New York, New York 10031}
\centerline{\it ${\rm ^2}$Department of Physics, New York University,
New York, New York 10003}
\centerline{January 30 1995}
\begin{abstract}
The interaction of an edge dislocation made of half
the superconducting plane with a magnetic interlayer vortex is considered
within the framework of the Lawrence-Doniach model with negative as well as
positive Josephson interlayer coupling. In the first case the binding
energy of the vortex and the dislocation has been calculated by employing
a variational procedure. The current distribution  around the bound vortex
turns
out to be asymmetric. In the second case the dislocation carries a
spontaneous magnetic half-vortex, whose binding energy with the
dislocation turns out to be infinite. The half-vortex energy has been
calculated by the same variational procedure. Implications of the possible
presence of such half-vortices for the properties of high temperature
superconductors are discussed. 
\end{abstract}
\end{titlepage}

\section{Introduction}
Many properties of high temperature superconducting (HTS) materials
are well understood within the framework of Lawrence-Doniach
model [1,2] assuming Josephson type coupling between the Cu-O
superconducting (SC) layers. Recently in [3] it has been proven
experimentally that the current-voltage and microwave characteristics
of these compounds
exhibit features specific to a stack of Josephson junctions
(JJ), each being 10$\div$15  \AA\ thick in the $c$-direction
[3]. To date, the LD model was widely employed
(see [4-7]) to describe the behavior of interlayer magnetic vortices
in an ideal layered structure.

In [8] Annett has suggested that the Josephson interlayer critical
current $J_c$ might be negative, if the BCS pairing mechanism
is confined within each layer and does not work between the layers.
In this case the ground SC state of the LD model is characterized
by  the phase shift $\pi$ of the order parameter in adjacent layers.
To distinguish such a state from the conventional one [1,2] and
following the terminology of [9], the former and the latter will
be named below as $\pi$-superconducting ground state ($\pi$-SC) and
 0-superconducting ground state (0-SC), respectively. It is worth noting that
several mechanisms may account for the presence of $\pi$-SC (or $J_c<0$).
As was suggested in [9], if the tunneling between two metals is controlled
by spin-flip processes, the Josephson critical current between these
metals should be negative. The other mechanism causing $J_c<0$ [10],
relies on the possibility that the tunneling could occur indirectly
through some intermediate state where Coulomb repulsion excludes a
double occupancy of such a state. We will not further discuss here
the extent to which these mechanisms are pertinent to the interlayer
tunneling in HTS compounds. Instead, we will consider how the existence
of $\pi$-SC, if any, may be recognized in specific features of vortex
dynamics.

The 0-SC and $\pi$-SC appear to be physically indistinguishable
in an ideal infinite crystal unless a screw or edge dislocation is
present [11]. Then in the case of the $\pi$-SC these
should carry a spontaneous half-vortex (HV) as a consequence of the
phase mismatch $\pi$, while traveling around the dislocation line.
Therefore an investigation of interaction between vortices and
dislocations can distinguish these two SC ground states.

In what follows we will consider the interaction of the interlayer
magnetic vortex (which has no normal core) with the edge linear
dislocation made of half a SC plane in the two cases (0-SC and
$\pi$-SC) of the LD model.\hfil\bigskip
\section{Model}
We will employ the LD model [1,2]. As was discussed in [2,6], because of
weak interlayer coupling, the constant amplitude approximation for the
in-plane order parameters can be adopted in order to describe interlayer
vortices. Thus, the LD free energy
functional becomes dependent on the phases $\Phi_n$ and the in-layer
components of the vector potential ${\bf A}_n$, $n$ labels the layers.
The edge dislocation can be modeled as a line $z=0,x=0$ so that for
$x>0$ the conducting half plane $n=0$ is missing (FIG.1). To incorporate
it into the LD model it is useful to separate the contribution of the
$n=0$ plane. Consequently, the free energy functional without the dislocation
can be written as [2]
\begin{eqnarray}
F&=&\int d^2{\bf
r}\,\bigg\{\sum_{n=1}^{+\infty}\big(K_n+U_n^+\big)+\sum_{n=-1}^{-\infty}
\big(K_n+U_n^-\big)
+{1\over 8\pi}\int dz\,(curl{\bf A})^2 \nonumber \\ &+&K_0+U_0\bigg\}
\quad,
\end{eqnarray}
where the notations are introduced as follows:
\begin{eqnarray}
K_n&\equiv &{1\over 8\pi}\left(\varphi_0\right)^2
{s\over \lambda_{ab}^2}\,\big({\bf \nabla}\Phi_n+
{2\pi \over \varphi_0}{\bf A}_n\big)^2\quad, \nonumber \\
U_n^{\pm}&\equiv &{\hbar\over 2e}J_c\big[1-\nu\cos{(\Phi_n-\Phi_{n\pm 1}-
\chi_{n,n\pm 1})}\big]\quad, \nonumber \\
U_0&\equiv &{\hbar\over 2e}J_c\big[2-\nu\cos{(\Phi_0-\Phi_{-1}-\chi_{0,-1}
)}-\nu \cos{(\Phi_1-\Phi_0-\chi_{1,0})}\big]\quad, \nonumber \\
\chi_{n,n+1}&\equiv &{2\pi\over \varphi_0}\int_{n}^{n+1}dz\,A_z \quad .
\nonumber
\end{eqnarray}
\noindent Here, the superconducting layers are infinitesimally thin and
separated by insulating layers of thickness $s$; the coordinates ${\bf
r}=(x,y)$ lie in the planes, and $z$ is directed perpendicular to the
layers;
$\varphi_0$ stands for the unit flux, $\lambda_{ab}$ is the in-plane
London penetration length, $J_c$ is the critical interlayer current;
${\bf \nabla}\equiv (\partial_x,\partial_y),{\bf A}\equiv (A_x,A_y)$.
One can see that only the last two terms of (1) contain $\Phi_0$.

The parameter $\nu$ is introduced here to distinguish between the two
possible states of the LD model: 0-SC ($\nu=1$) [2] and $\pi$-SC (
$\nu=-1$) [8]. In the format of (1) these two
states can be converted into each other by the phase shift $\Phi_n
\rightarrow\Phi_n+\pi n$. However, the presence of a dislocation
makes such a conversion impossible.

To describe a dislocation in the $n=0$ layer (FIG.1) one should replace
the Josephson (potential) energy $U_0$ by the following expression:
\begin{equation}
	U_{0d}={\hbar\over 2e}\,J_c\,\left\{
		\begin{array}{ll}
		2-\nu\cos{(\Phi_0-\Phi_{-1}-\chi_{0,-1})}-\nu
		\cos{(\Phi_1-\Phi_0-\chi_{1,0})}\,, & x<0 \\
		1-\nu\cos{(\Phi_{-1}-\Phi_{1}-\chi_{-1,1})}\,, &x>0\,.
		\end{array}
		\right.
\end {equation}
\noindent Equation~(2) signifies that to the left of the dislocation
line ($x<0$), where the plane $n=0$ exists, the interaction with the
layers $n=1$ above and $n=-1$ below is the same as given by (1). However
, to the right of this line ($x>0$) the half-plane $n=0$ is missing and,
therefore the layer $n=1$ appears to be coupled to the $n=-1$ with the
same current $J_c$. In our model we ignore any role of the dislocation
 core, because
its size is much smaller than the London or Josephson lengths.

If a vortex resides exactly at and along the dislocation line, the lines
of current should possess the symmetry with respect to the mirror reflection in
the plane $z=0$. Therefore, no current flows along the
half-plane $n=0$. It implies that the solution for such a vortex in the
gauge $A_z=0$ obeys the conditions:
\begin{equation}
	{\bf \nabla}\Phi_0+{2\pi \over \varphi_0}{\bf A}_0=0\quad,\quad
	\Phi_{1}=-\Phi_{-1} \quad .
\end{equation}
Making use of these conditions, one finds that the potential energy
$U_0$ in (2) should be replaced by
\begin{eqnarray}
	U_{0d}&=&{\hbar\over 2e}J_c\left\{
		\begin{array}{ll}
		2\,(1-\nu\cos{{\Phi_s\over 2}})\,, & x<0 \\
		1-\nu\cos{\Phi_s}\,, & x>0
		\end{array}
  		\right.\quad ;\\
\Phi_s({\bf r})&\equiv &\Phi_1({\bf r})-\Phi_{-1}({\bf r})
\quad .
\nonumber
\end{eqnarray}
Assuming the system is in the 0-SC state ($\nu=1$), the lowest energy
solution can be achieved for $\Phi_s=0$. If, however, $\nu=-1$ ($\pi$-SC),
the lowest energy corresponds to the asymptotics
\begin{equation}
\Phi_s(-\infty)=\pm2\pi\quad,\quad \Phi_s(+\infty)=\pm\pi
\end{equation}
These account for a half-flux contained in the $z=0$ plane. Indeed,
employing the expression [12]
\begin{equation}
	\varphi={\varphi_0\over
2\pi}\,\big(\Phi_s(+\infty)-\Phi_s(-\infty)\big)\quad .
\end{equation}
for the total Josephson flux $\varphi$, one finds that
$\varphi=\mp\varphi_0/2$. It is worth noting that imposing the condition
$\Phi_s(+\infty)=\Phi_s(-\infty)$ instead of (5) in order to remove the
HV
makes (4) acquire a contribution proportional to the area either of the
half-plane $n=0$ or the missing part of this plane.
Therefore, we conclude that a single edge dislocation
in $\pi$-SC binds the HV with an infinite energy.

To find the energy of the HV in $\pi$-SC as well as that of the integer
vortex in 0-SC trapped at the dislocation line, we will employ the
continuum limit for all the layers but the three central ones
$n=-1,0,1$. It means that the potential energy (4) is retained and the
summation in (1) is replaced by integration over two regions $z>+0$ and
$z<-0$. We expand
\begin{equation}
	\cos{(\Phi_n-\Phi_{n+1}-\chi_{n,n+1})}\approx 1-{s^2\over 2}
	\left(\partial_z\Phi+{2\pi\over \varphi_0}A_z\right)^2
\end{equation}
for the 0-SC and the same for the $\pi$-SC, with the replacement $\Phi_n
\rightarrow\Phi_n+\pi n$ made.
  The continuum approximation turns out to be valid, as discussed
in [2,4], outside of the junction where the vortex resides, as long as the
relation $\lambda_{ab}/s\gg 1$ holds.
 Therefore, in the following consideration the thickness
of the central junction made of two infinite layers $n=1,n=-1$, with the
half-plane $n=0$ intervening in between, is
infinitesimal.
 As a result, we arrive at the
functional (1) written as
\begin{eqnarray}
F&=&{1\over 8\pi}\bigg({\varphi_0\over 2\pi}\bigg)^2\int_{
                z<0,z>0
		}
		d^2{\bf r}\,dz\,\bigg[\,{1\over \lambda_{ab}^2}\Big({\bf \nabla}\Phi+
	{2\pi\over \varphi_0}{\bf A}\Big)^2+{1\over \lambda_c^2}\Big(\partial_z
	\Phi+{2\pi \over \varphi_0}A_z\Big)^2\bigg] \nonumber \\
	&+&{1\over 8\pi}\int d^2{\bf r}\,dz\,\Big(curl{\bf A}\Big)^2+\int d^2
	{\bf r}\,U_{0d}\quad, \\
\lambda_c^{-2}&\equiv &{h\over e}\,\bigg({2\pi \over \varphi_0}
\bigg)^2J_c\,s\quad, \nonumber
\end{eqnarray}
where the conditions (3) and (4), which derives from (3), are taken into
account.

Variation of (8) allows one to obtain the linearized bulk equations (the
problem is homogeneous in the $y$-direction and the gauge is $A_z=0$) as
follows:
\begin{equation}
	{2\pi\over \varphi_0}\,\partial_z^2A_x={1\over \lambda_{ab}^2}
\left(\partial_x\Phi+{2\pi\over \varphi_0}A_x\right)\quad ,
\end{equation}
and
\begin{equation}
	{2\pi\over\varphi_0}\,\partial_{xz}^2A_x=-{1\over \lambda_c^2}\,
	\partial_z\Phi\quad .
\end{equation}
The solution for the gauge invariant phase $\phi=\Phi+{2\pi\over
\varphi_0}\eta$, where the magnetic potential $A_x=\partial_x\eta$ is
introduced, can be easily found from Eqs.~(9)-(10), if one Fourier transforms
along the $x$-axis, yielding
\begin{eqnarray}
	\phi_q(z)=\left\{\begin{array}{ll}
		\phi_q(+0)\,e^{-Q_qz}\,, & z>0 \\
		\phi_q(-0)\,e^{Q_qz}\,, &z<0
		\end{array}
		\right.\quad ;\\
	Q_q\equiv \lambda_{ab}^{-1}(1+\lambda_c^2q^2)\quad,\quad
	\phi_q(+0)+\phi_q(-0)=0\quad . \nonumber
\end{eqnarray}
Here, $\phi_q(+0),\phi_q(-0)$ stand for the Fourier components of the
gauge invariant phase at the upper and lower edges $z=+0,z=-0$,
respectively, of the
central junction. Integrating by parts with respect to $z$ in (8) and
making use of (9)-(10) one can express the free energy in terms of the
surface integration only. Taking into account that the magnetic
potential $\eta$ has no jump at $z=0$ and that the phase does have a
jump, one obtains the line energy (free energy per unit length in the
$y$-direction) as
\begin{eqnarray}
\varepsilon & = & {1\over 4\pi}\bigg({\varphi_0\over 2\pi}\bigg)^2
	\bigg\{\,{1\over 4\lambda_{ab}}\sum_q{q^2\over \sqrt{1+\lambda_c^2
	q^2}}|\phi_{sq}|^2 \nonumber \\
	& + & {1\over
s\lambda_c^2}\,\Big[\,\int_{-\infty}^0dx\,2\,(1-\nu\cos{{\phi_s\over 2}})+
	\int_0^{+\infty}dx\,(1-\nu\cos\phi_s)\Big]\bigg\}\quad ,
\\
\phi_{sq}&\equiv&\phi_q(+0)-\phi_q(-0)\quad . \nonumber
\end{eqnarray}
It is worth noting that because of the limit $\lambda_{ab}/s\gg1$, the jump of
the gauge invariant phase $\phi_{sq}$ is taken to be equal to the
jump $\Phi_s$ (4) of the phase $\Phi$. In the next section we will
utilize (12) to calculate the energy of an integer vortex at the
dislocation line (0-SC case), and show that this energy is less than
that of the vortex far from the dislocation, implying a binding of the
vortex to the dislocation line.
\section{Energy of an integer vortex bound to a dislocation}
\noindent If the penetration lengths $\lambda_{ab}$ were compared to
$\lambda_c$, one would have converted (12) to the conventional sine-Gordon
functional for the JJ [12] making the replacement
\mbox{$\sqrt{1+\lambda_c^2q^2}
\approx1$}, and \mbox{$\sum_qq^2|\phi_{sq}|^2\rightarrow\int dx(\partial_x
\phi_s)^2$} in (12). Consequently, in the $\pi$-SC the HV solution would
be obtained. However, the anisotropy in our case will play an essential role,
implying that (12) cannot be represented in a closed form in the
physical space of one coordinate $x$: an infinite number of the gradient
 terms should be
retained. To obtain a rigorous upper bound on the vortex energy in the
0-SC we will employ the variational trial function:
\begin{eqnarray}
	\phi_s(x)&=&\left\{\begin{array}{ll}
		C_-e^{\alpha kx}+B_-e^{{x\over \lambda_c}}, & x<0\\
		2\pi -C_+e^{-kx}-B_+e^{-{x\over \lambda_c}}, &x>0
		\end{array}\right.\quad ; \\
	C_+&=&{1\over 1+\alpha}\left[\,2\pi -(1-{1\over \lambda_ck})B_+-
	(1+{1\over \lambda_ck})B_-\right]\, , \nonumber \\
	C_-&=&{\alpha\over 1+\alpha}\left[\,2\pi -(1+{1\over \lambda_ck})B_+-
	(1-{1\over \lambda_ck})B_-\right]\,. \nonumber
\end{eqnarray}
Here, $\alpha,k,B_{\pm}$ are variational parameters. It can be seen
that (13) describes a function which is continuous with its first
derivative. This form of the trial function corresponds to the magnetic
field pointing in the $y$-direction, the total flux being $\varphi_0$
(see (6)). The term proportional to $\exp{(\pm x/ \lambda_c)}$ was
included to account for the fact that at large distances from the vortex
 the solution of (9-10) decays with such exponents along the $x$-axis.
However, it can be shown that all the terms proportional to $B$\/'s
produce a contribution into the energy as small as $s/\lambda_{ab}\ll1$.
Thus, in what follows, we put $B_{\pm}=0$ in (13).

Fourier transforming (13) and substituting the result into (12), we
arrive at the line energy of the vortex residing at the dislocation as
\begin{eqnarray}
	\varepsilon_{1d}&=&\bigg({\varphi_0\over 4 \pi}\bigg)^2
	{1\over \lambda_{ab}\lambda_c}\left[\,\ln{(k\lambda_c)}+\ln{2}-
	{\ln{\alpha}\over \alpha^2-1}+I(\alpha){\lambda_{ab}\over
	s}{1\over k\lambda_c}+O\left({s\over \lambda_{ab}}\right)
	\right] \\
	I(\alpha)&\equiv&{2\over \pi}\left[\,\int _0^{{\alpha\pi\over 1+
	\alpha}}du\,{\sin{u}^2\over u}+{2\over \alpha}\int_0^{{\pi\over
	2(1+\alpha)}}du\,{\sin{u}^2\over u}\right]\quad . \nonumber
\end{eqnarray}
Minimization of (14) with respect to $k$ gives
\begin{equation}
	k\lambda_c=I(\alpha){\lambda_{ab}\over s}\gg1\quad .
\end{equation}
The energy (13) then can be written as
\begin{equation}
	\varepsilon_{1d}=\bigg({\varphi_0\over 4 \pi}\bigg)^2
	{1\over \lambda_{ab} \lambda_c}\left[\,\ln{{\lambda_{ab}\over s}}
	+1+\ln{2}-{\ln{\alpha}\over \alpha^2-1}+\ln{I(\alpha)}
	+O\left({s\over \lambda_{ab}}\right)
	\right]\quad .
\end{equation}
This can be minimized numerically with respect to $\alpha$, producing
\begin{equation}
	\varepsilon_{1d}=\bigg({\varphi_0\over 4 \pi}\bigg)^2
	{1\over \lambda_{ab} \lambda_c}\left[\,\ln{{\lambda_{ab}\over s}}
	+1.05+O\left({s\over \lambda_{ab}}\right)\right]
\end{equation}
at $\alpha=0.85$, suggesting an asymmetric distribution of currents
about the dislocation line.

To find the vortex energy far from the dislocation one can employ the
same variational function (13), where the symmetric solution
($\alpha=1$) should be looked for, and the potential energy ought to be
replaced by the second line of (4) in a whole space as
\begin{equation}
	U_0={\hbar\over 2e}J_c\,(1-\nu\cos{\phi_s})
		\quad , \quad -\infty<x<+\infty \quad .
\end{equation}
Then, the rigorous upper bound for the integer vortex line energy far
from the dislocation is
\begin{equation}
	\varepsilon_{1}=\bigg({\varphi_0\over 4 \pi}\bigg)^2
	{1\over \lambda_{ab} \lambda_c}\left[\,\ln{{\lambda_{ab}\over s}}
	+1.24+O\left({s\over \lambda_{ab}}\right)\right]\quad .
\end{equation}
A comparison of (17) and (19) shows that the latter is higher,
suggesting that the vortex binds to the dislocation. It is worth noting
that the exact evaluation of the line energy of an interlayer vortex
 in the ideal layered crystal obtained in [4] gives
\begin{equation}
	\varepsilon_{1}=\bigg({\varphi_0\over 4 \pi}\bigg)^2
	{1\over \lambda_{ab} \lambda_c}\left[\,\ln{{\lambda_{ab}\over s}}
	+1.12+O\left({s\over \lambda_{ab}}\right)\right]\quad ,
\end{equation}
which appears to be higher than the energy (17) of the integer vortex
bound to the dislocation as well. Comparing (19) and (20), we can say
that the variational procedure suggested above gives a correct result in
the leading logarithmic ($\ln{(\lambda_{ab}/s)}\gg1$) approximation. At
the same time, the second term in brackets (19) turns out to be
overestimated by 10\% in comparison with that in (20). Therefore,as soon
as the bound vortex line energy (17) contains the same $\ln{(\lambda_{ab}/c)}$,
 we expect
that the binding energy is considerably underestimated. This, however,
can be improved by introducing additional terms into the variational
ansatz (13).
\section{The line energy of the half-vortex in the $\pi$-SC.}
In the case [8] of the $\pi$-SC ($\nu=-1$) the lowest energy solution
obeys the asymptotic behavior (5). Therefore, the trial function can be
taken in the form
\begin{equation}
	\phi_s(x)=\left\{\begin{array}{ll}
		2\pi-{\pi\over 1+\alpha}e^{\alpha kx}, &x<0\quad , \\
		\pi+{\pi\alpha\over 1+\alpha}e^{-kx}, & x>0\quad ,
		\end{array}\right.
\end{equation}
where the tail $\sim \exp{(\pm x/\lambda_c)}$ has been omitted, as discussed in
Section~3. Calculations similar to those performed above for the case
$\nu=1$ yield the solution for the variational parameters as
\begin{equation}
	\lambda_ck=1.09\,{\lambda_{ab}\over s}\quad,\quad \alpha=0.72\quad
,
\end{equation}
and the half-vortex line energy
\begin{equation}
	\varepsilon_{1/2}={1\over 4}\,\bigg({\varphi_0\over 4 \pi}\bigg)^2
	{1\over \lambda_{ab} \lambda_c}\left[\,\ln{{\lambda_{ab}\over s}}
	+1.21+O({s\over \lambda_{ab}})\right] \quad .
\end{equation}
A comparison of this result with (19) shows that in the leading
logarithmic approximation $\ln{(\lambda_{ab}/s)}\gg1$ the line energy of the
half-vortex is 1/4 that of the integer vortex, as one would expect from
estimates [2,4] where the unit flux $\varphi_0$ is replaced by its half
$\pm\varphi_0/2$.
\section{Discussion.}
As was emphasized above, an integer vortex turns out to be the same in the
both states ($\nu=1,\nu=-1$) of the LD model. It does not allow
 one to distinguish these
states by investigating the behavior of such a vortex unless it
encounters a dislocation. In the 0-SC, where the SC order parameter in
the ground state has the same sign in all the layers, the edge
dislocation binds the vortex with the line energy not less than
\begin{equation}
	\varepsilon_b=0.07\,\bigg({\varphi_0\over 4 \pi}\bigg)^2{1\over \lambda_{ab}
	\lambda_c}\quad .
\end{equation}
This estimate is obtained by a comparison of the exact result (20)
obtained in [4] for an
interlayer vortex energy in an ideal layered crystal with the
variational energy (17) for the bound integer vortex.

In the $\pi$-SC, where the SC order parameters in the ground state have
opposite signs in adjacent layers [8], the spontaneous half-vortex
exists inherently at the dislocation line. Therefore, an integer vortex
approaching this dislocation will either be repelled or attracted to
it, depending on the mutual orientation of the two vortices. If their
moments are opposite, the integer vortex will recombine with the
half-vortex. As a result of such a process, the integer vortex
disappears, and the half-vortex switches orientation. It means that the
energy of the former will be released in some way.

Given these two different types of interaction of free vortices with a linear
edge dislocation, the two possible states of the LD model can be
distinguished by investigating the dynamics of vortices in a single HTS
crystal with a density of interlayer edge dislocations less than the
reciprocal area ($1/ \lambda_{ab} \lambda_c$) occupied by one vortex.
Under this condition the pairs of dislocations and vortices bound to them
 can be
considered independently of each other, and our results will apply. Thus, in
the case
of the 0-SC these dislocations should play a role of weak pinning
centers. Therefore, after imposing and then removing an external
magnetic field along the layers some residual magnetisation might be observed,
because
 of the vortices
trapped by the dislocations. Effects of thermal
fluctuations and weak vortex-vortex interactions will cause this
magnetisation to decay slowly with the typical time determined by the
line binding energy (24).

As it can be shown, in the $\pi$-SC the ground state of an array
 of half-vortices attached to
dislocations in a single crystal will be antiferromagnetic. Hence, no
macroscopic
magnetisation is expected unless an external field along the layers is imposed.
Introducing external vortices, with their density being twice less than that
of the dislocations, will result in them recombining with that part of
the half-vortices which have the orientation opposite to the external
field. As a result, a residual magnetisation of the sample will occur.
This magnetisation, which was originally due to the integer vortices,
 is now produced by the half-vortices ordered
ferromagnetically. With the external field removed, such a configuration is not
the lowest in energy, and
these half-vortices should re-orient themselves to return to their former
antiferromagnetic ordering. However, to do so the half-vortex should
emit an integer vortex which, then, is to be expelled from the sample.
 The energy required for such a process is that of the integer vortex
, and, as comparison of (19),(20) with (24) shows, it is bigger than the
 binding energy (24) in the 0-SC by the factor of
$\ln{(\lambda_{ab}/s)}\gg1$. Therefore, in the $\pi$-SC the residual
magnetisation, which is controlled by the dislocation density, should
decay with a typical time constant much bigger than in the 0-SC.

The other implication of the $\pi$-SC was discussed in [11]. If a SC
grain is in the $\pi$-SC and contains an odd number of dislocations, it
should carry a net magnetisation $\pm\varphi_0/2$ because of the
uncompensated half-flux. An array of such grains coupled by means of the
magnetic field only would exhibit paramagnetic response. In this respect
the paramagnetism of the HTS granular materials [13] might be considered
as an indication of the existence of the $\pi$-SC. The recent
observation [14] of the half-flux trapped inside a HTS ring could also
indicate that the grains in this ring are in the $\pi$-SC, and that they
are joined in such a way that the Cu-O layers form a screw-like
structure, while traveling around the ring. In this case such a ring
characterized by the structural chirality should develop a spontaneous
half-flux.

In conclusion, we employed a variational approach to show that a single
interlayer edge dislocation binds an integer interlayer (coreless) vortex,
within the framework of the LD model. The current distribution around
such a bound vortex turns out to be asymmetric. Regarding the suggestion
[8] of the $\pi$-type interlayer Josephson coupling, we have shown that
half-vortex is attached to the dislocation line, and calculated the
half-vortex line energy. The physical consequences of such a ground
state were discussed.

We are grateful to John R. Kirtley for useful discussions of possible
methods for experimental detection of half-vortices.

\vfill\eject
\noindent {\Large\bf Figure Caption}\\ \break
\noindent Fig.1 Schematics of the edge dislocation made of half of the
conducting
plane $n=0\,(\,z=0, x<0,-\infty<y<+\infty\,)$ which is inserted between
the two layers $n=\pm1$. The axis $z$ is perpendicular to the layers
far from the dislocation core, which location is indicated by the vertical
bar.
\end{document}